\documentclass[prl,twocolumn,showpacs,superscriptaddress,floats]{revtex4-1}

\usepackage{amsmath,amssymb}
\usepackage{graphicx}

\newcommand{\kv}{\vec{k}}

\begin{document}

\title{Extremely Correlated Fermi Liquid Description of Normal State ARPES
in Cuprates}

\author{G.-H.~Gweon}
\email{gweon@ucsc.edu}
\affiliation{Physics Department, University of California, Santa Cruz, CA,
95064, USA}

\author{B.~S.~Shastry}
\email{sriram@physics.ucsc.edu}
\affiliation{Physics Department, University of California, Santa Cruz, CA,
95064, USA}

\author{G.~D.~Gu}
\affiliation{Condensed Matter Physics and Materials Science Department,
Brookhaven National Laboratory, Upton, New York 11973, USA}

\date{\today}

\begin{abstract}
The normal state single particle spectral function of the high temperature
superconducting cuprates, measured by the angle resolved photoelectron
spectroscopy (ARPES), has been considered both anomalous and crucial to
understand.  Here, we report an unprecedented success of the new
Extremely Correlated Fermi Liquid theory by Shastry to describe
both laser and conventional synchrotron ARPES data (nodal cut at optimal
doping) on Bi$_2$Sr$_2$CaCu$_2$O$_{8+\delta}$ and synchrotron data on
La$_{1.85}$Sr$_{0.15}$CuO$_4$.  {\em It fits all data sets with the same
physical parameter values}, satisfies the particle sum rule and
successfully addresses two widely discussed ``kink'' anomalies in the
dispersion.
\end{abstract}


\maketitle

Angle resolved photo-electron spectroscopy (ARPES) was  the first probe to
provide a detailed view of the anomalous nature of high temperature cuprate
superconductors, discovering  unexpectedly broad spectra with intense and
asymmetric tails that have remained an enduring mystery for the last two
decades.  Conventional data taken with high energy ($\gtrsim 15$ eV)
photons from synchrotron light sources have recently been supplemented with
laser ARPES data \cite{koralek_laser_2006, zhang_identification_2008} from
lower energy (6 or 7 eV) sources.  The latter  show considerably sharper
features near the Fermi energy.  A drastic possibility to account for this
distinction is that the sudden approximation could break down for the
smaller photon energies used in laser ARPES \footnote{In the sudden
approximation, the ARPES intensity $I(\kv,\omega) = M(\kv) A (\kv,\omega)
f(\omega)$.  Here, $\kv$ is the momentum and $-\omega$ is the energy of the
final $N-1$ state (as we set $\hbar = 1$ in this Letter), where $N$ is the
number of electrons in the initial state.  $A(\kv,\omega) = \Im m\
G(\kv,\omega- i 0^+) / \pi$ is the single particle spectral function,
$M(\kv)$ is the dipole matrix element, constant for an EDC, and $f(\omega)$
is the Fermi-Dirac function.  At the chemical potential $\omega \equiv 0$.}

An important un-answered question is whether the results of the two
spectroscopies could be reconciled in a single  theoretical framework
that does not abandon the sudden approximation.  More broadly, can we
understand the wide variety of observed lines shapes in a theoretical
framework with a sound microscopic basis and a {\em single} set of
parameters?

In this Letter, we confront a recent theory of Extremely Correlated Fermi
Liquids (ECFL) proposed by Shastry \cite{shastry_extremely_2011} with the
above challenge.  The new formalism is complex and requires considerable
further  effort to yield numerical results in low dimensions.  In the limit
of high enough dimensions, however, a remarkably simple expression for the
Green's function emerges; it is significantly different from the standard
Fermi Liquid Dyson form, while satisfying the usual sum rules.  We use this
simple version of ECFL  Green's function in this Letter, motivated by the
attractive spectral shapes produced with very few parameters
\cite{shastry_extremely_2011}.  {\em In this Letter we show that already
the simplest version of the ECFL theory, with very few parameters, is
successful to an unprecedented extent in detailed fitting of a wide variety
of normal state cuprate ARPES line shapes.} Interesting predictions are
made for the higher temperature spectral line skew.

Our focus in this Letter is on the data of optimally doped
Bi$_2$Sr$_2$CaCu$_2$O$_{8+\delta}$ (Bi2212) and
La$_{1.85}$Sr$_{0.15}$CuO$_4$ (LSCO) superconductors in the normal state,
taken with $\kv$ along the nodal direction connecting $(0,0)$ to
$(\pi/a,\pi/a)$.  Most of the data is taken from the published literature,
while some original data are also presented (Bi2212 data in Figs.\@
\ref{fit-ours},\ref{kinks}).  Our sample is an optimally doped Bi2212
($T_c$ = 91 K), grown by the floating zone method at the Brookhaven
National Laboratory (BNL), and was measured at the Stanford Synchrotron
Radiation Lightsource (SSRL) beam line 5-4 using 25 eV photons.  The
resolutions are 15 meV (energy) and 0.3$^\circ$ (angle).

{\bf Line shape model:}\ \ The ECFL spectral function is given as a product
of an auxiliary Fermi Liquid (aux-FL) spectral function
$A_{FL}(\vec{k},\omega)$ and a second frequency dependent ``caparison''
factor \cite{shastry_extremely_2011, shastry_anatomy_2011}:
\begin{eqnarray}
	A(\kv,\omega) &=& A_{FL} (\kv,\omega) \left( 1 - \frac{n}{2} +
	\frac{n^2}{4} \cdot \frac{\xi_{\kv} - \omega}{\Delta_0} \right)_+
	\label{ecfl-A}
\end{eqnarray}
where $n$ is the number of electrons per CuO$_2$ unit cell, $(X)_+ \equiv
\max (X,0)$, $\xi_{\kv} = \left(1-\frac{n}{2}\right)\, \varepsilon({\kv})
$, where $\varepsilon({\kv})$ is the bare one-electron band dispersion (see
later).  Here, $A_{FL}(\kv,\omega) = \frac{1}{\pi} \cdot \Im m
\frac{1}{\omega - \xi_{\kv} - \Phi (\omega)}$ with
\begin{eqnarray}
    \Im m\ \Phi (\omega) &=& \frac{\omega^2+\tau^2}{\Omega_0} \exp \left(-
	\frac{\omega^2 + \tau^2}{\omega_0^2}\right) + \eta
	\label{afl-ImPhi}
\end{eqnarray}
where $\tau = \pi k_B T$, $T$ is the temperature, and $\omega$ is to be understood as $\omega - i0^+$.
Here, $\omega_0$ is the
aux-FL energy scale (i.e.\@ high $\omega$ cutoff), and $\Omega_0$ governs
the lifetime, and, by causality, the quasi-particle weight (i.e.\@ the wave
function renormalization) of the aux-FL,
$Z_{FL} = (1+\frac{\omega_0}{\sqrt{\pi}\Omega_0})^{-1}$, as identified
from $\Re e \Phi $\footnote{$ \Re e\ \Phi (\omega) = - \frac{1} {\sqrt{\pi}
\Omega_0} \exp \left(-\frac{\tau^2}{\omega_0^2}\right)\, \left[ \omega_0
\omega - 2 (\omega^2+\tau^2) D\left(\frac{\omega}{\omega_0}\right) \right]$
where $D(x) = \frac{\sqrt{\pi}}{2} e^{-x^2}\mathrm{erfi}(x)$ is the Dawson
function. $C_\Phi = 1/(\pi \Omega_0)$ and $\omega_c = C_\Phi \omega_0^2$ where $C_\Phi$ and $\omega_c$ are parameters defined in Ref.~[\onlinecite{shastry_extremely_2011}].}.

The ECFL energy scale $\Delta_0$ measures the ``average intrinsic
in-elasticity'' of the aux-FL.  It is given \cite{shastry_extremely_2011}
as
\begin{eqnarray}
	\Delta_0 &=& \int_{-\infty}^{\infty} d\omega f(\omega) \langle A_{FL}
	(\kv,\omega) (\xi_{\kv} - \omega) \rangle_{BZ} \label{ecfl-Delta0}
\end{eqnarray}
where $\langle \cdot \rangle_{BZ}$ denotes  averaging  over the first
Brillouin zone.

The parameters that enter this description are now listed.  The  ``primary
parameters''  defining the ECFL fit  consist of the dispersion  $\xi_{\kv}$
taken from band theory, the density  $n$, temperature $T$, and the aux-FL
parameters $\Delta_0$, $\omega_0$, $Z_{FL}$, $\Omega_0$.   Of the last four
parameters,  only two are free parameters.  For instance, $\omega_0$ and
$Z_{FL}$ can be taken as free parameters, and $\Omega_0$ and $\Delta_0$ can
be calculated using the equation for $Z_{FL}$ and Eq.\@ \ref{ecfl-Delta0},
respectively. 

\begin{figure}
\begin{center}
	\includegraphics[width=0.95\columnwidth]{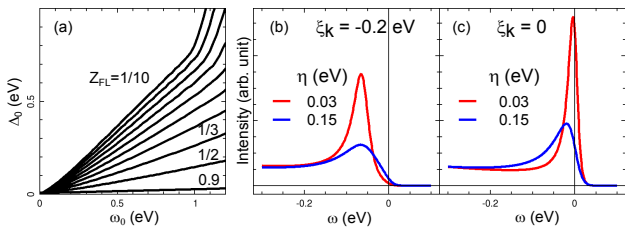}
	\caption{(a) $\Delta_0$ as a function of  $\omega_0$ for various
	$Z_{FL}$.  Other primary ECFL parameter values are $n=0.85$, $T = 100 $
	K, and $\xi_{\kv}$ as described in the text.  A small $\eta$ value,
	0.010 eV, was used for this plot, which is used as a ``lookup table''
	during the fit.  (b,c) Examples of the spectral function calculated
	with different values of the {\em effective sample quality parameter}
	$\eta$. See the caption of the next figure for parameter values used.
	The instrumental energy broadening of 10 meV (FWHM) is included.
	\label{Delta0_eta}}
\end{center}
\end{figure}

The parameter $\eta$  in Eq.~(\ref{afl-ImPhi}) is an additional ``secondary
parameter'' \footnote{By secondary, we mean ``intrinsic (i.e.\@ part of
$A(\kv,\omega)$) but dependent on the effective sample quality.''} with
respect to the  ECFL theory \cite{shastry_extremely_2011}. Its origin is in
impurity scattering as argued in \cite{abrahams_what_2000}, and
additionally, in scattering with surface imperfections.  Our fits determine
$\eta \approx 0.03$ eV for laser ARPES and $\eta \approx 0.15$ eV for
conventional ARPES.  Greater penetration depth in laser ARPES suggests that
it should be {\em less sensitive to surface imperfections}, thereby
yielding a smaller $\eta$.  We therefore propose that this parameter
summarizes the {\em effective sample quality} in different experiments.
The difference in line shapes arising from these values of $\eta$ is
demonstrated in Figs.\@ \ref{Delta0_eta}(b,c).

Our strategy is to fix a common set of  intrinsic parameters  for all the
materials, and allow $\eta$ to be determined separately for each class of
data.  The most time consuming part is the calculation of $\Delta_0$, the
results of which are summarized in Fig.\@ \ref{Delta0_eta}(a).

In our line shape analysis (1) we first set $n = 0.85$, corresponding to
the optimal doping. (2) Here $\xi_{\kv}$ is taken to be the un-renormalized
band dispersion, taken from the literature \cite{markiewicz_one-band_2005},
and then scaled to fit the observed occupied band width, 1.5 eV, of the
Bi2212 ARPES result \cite{meevasana_hierarchy_2007}\footnote{Also, a minor
$k$-shift of the theory was necessary ($\sim 0.02$ \AA) to match the
slightly different $k_F$ values reported in different experiments}. (3) We
choose $Z_{FL} = 1/3$, to account for the dispersion renormalization due to
the high energy kink \cite{graf_universal_2007, meevasana_hierarchy_2007},
which in this theory is caused by the energy scale $\omega_0$ (cf.\@ Fig.\@
\ref{kinks}).  (4) Finally, in all simulations, we include the finite
energy resolution effect and the finite angle resolution effect as a
combined Gaussian broadening (10 meV FWHM for laser ARPES and 25 meV FWHM
for conventional ARPES) in energy \footnote{The minor effect of the finite
angular resolution, significant only for the conventional ARPES data, can
be modelled well by an energy broadening.}.

\begin{figure}
\begin{center}
	\includegraphics[width=0.75\columnwidth]{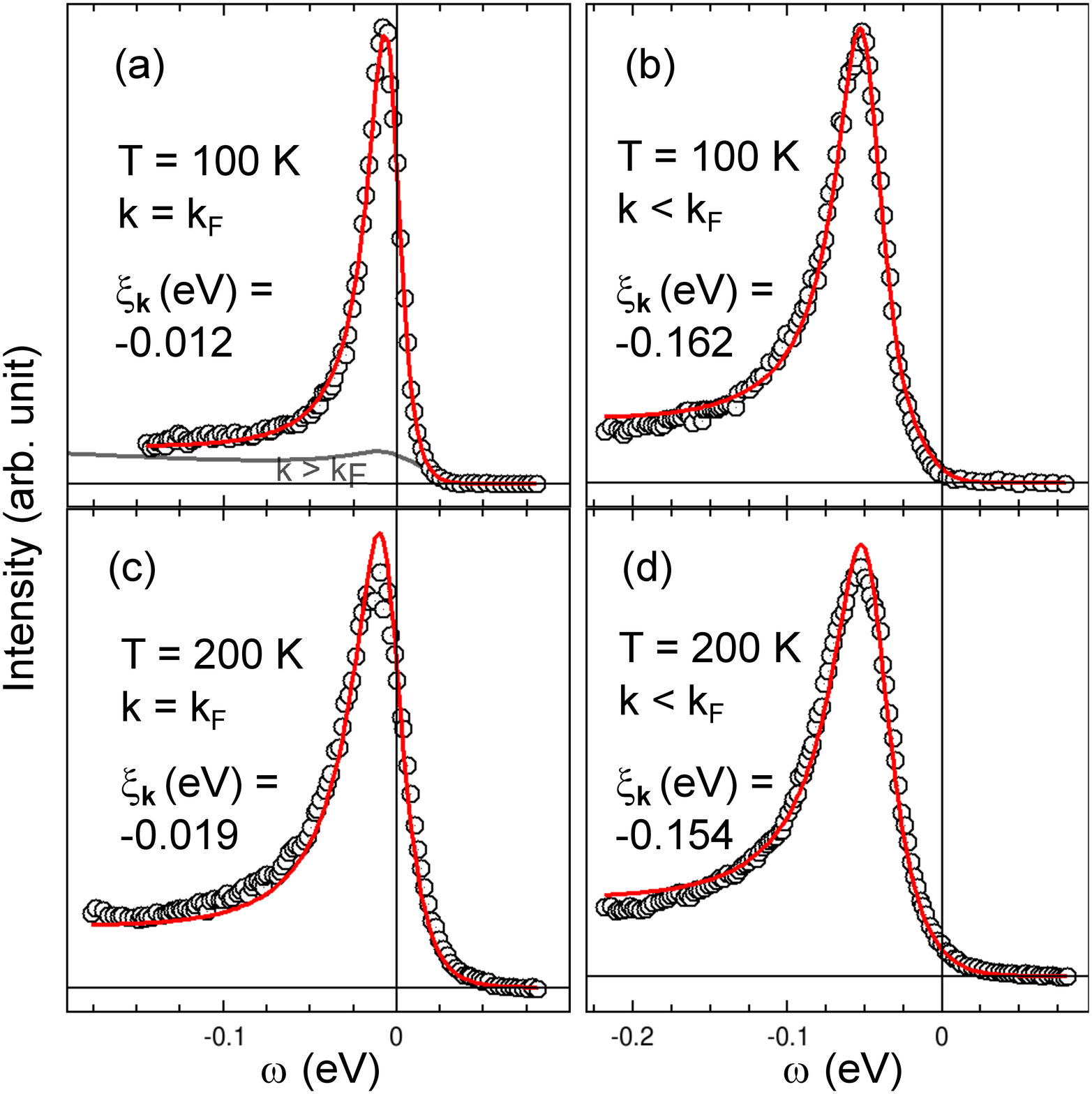}
	\caption{Laser ARPES data (symbols, Bi2212) from Ref.\@
	\onlinecite{casey_accurate_2008} fit with the ECFL line shape (red
	lines).  The free parameters of the fit were $\omega_0$ (0.5 eV),
	$\eta$ (0.032 eV), and $\xi_{\kv}$ (shown).  Fixed parameters: $n$
	(0.85), $Z_{FL}$ (1/3).  Derived parameters: $\Delta_0$ (0.12 eV),
	$\Omega_0$ (0.14 eV). Other than $\eta$ and $\xi_{\kv}$, the same
	parameters are used elsewhere in the Letter. In (a), the gray line
	corresponds to the theoretical curve with $\xi_{\kv} = 0.15$ eV\@.
	\label{fit-casey}}
\end{center}
\end{figure}

{\bf Line shape fit for laser ARPES:}\ \ Fig\@. \ref{fit-casey} shows the
fit of the laser ARPES data with the ECFL line shape.  These fits were made
using a procedure that is somewhat more restrictive than that in the recent
work of Casey  and Anderson
\cite{casey_accurate_2008,anderson_strange_2006},
using the equations of Doniach-Sunjic, Anderson-Yuval, and Nozieres-de
Dominicis
\cite{doniach_many-electron_1970, *nozieres_singularities_1969, *yuval_exact_1970} (CADS):  we are using global, rather than
per-spectrum, fit parameters.  However, our fit is somewhat less restricted
than other fits shown in this Letter: here we allow a small variation of
$\xi_{\kv}$ as in Ref.\@ \onlinecite{casey_accurate_2008}.  We find an
excellent fit quality, at least comparable to  CADS
\cite{casey_accurate_2008}.
The gray line in panel (a) shows our calculation for
$k > k_F$.  Our  expectation is that, were the data for $k > k_F$
available, we would find a reasonable fit in this $k$ region as well
\cite{jacob_koralek_laser_2006}, as for other data sets below.

\begin{figure}
\begin{center}
	\includegraphics[width=0.95\columnwidth]{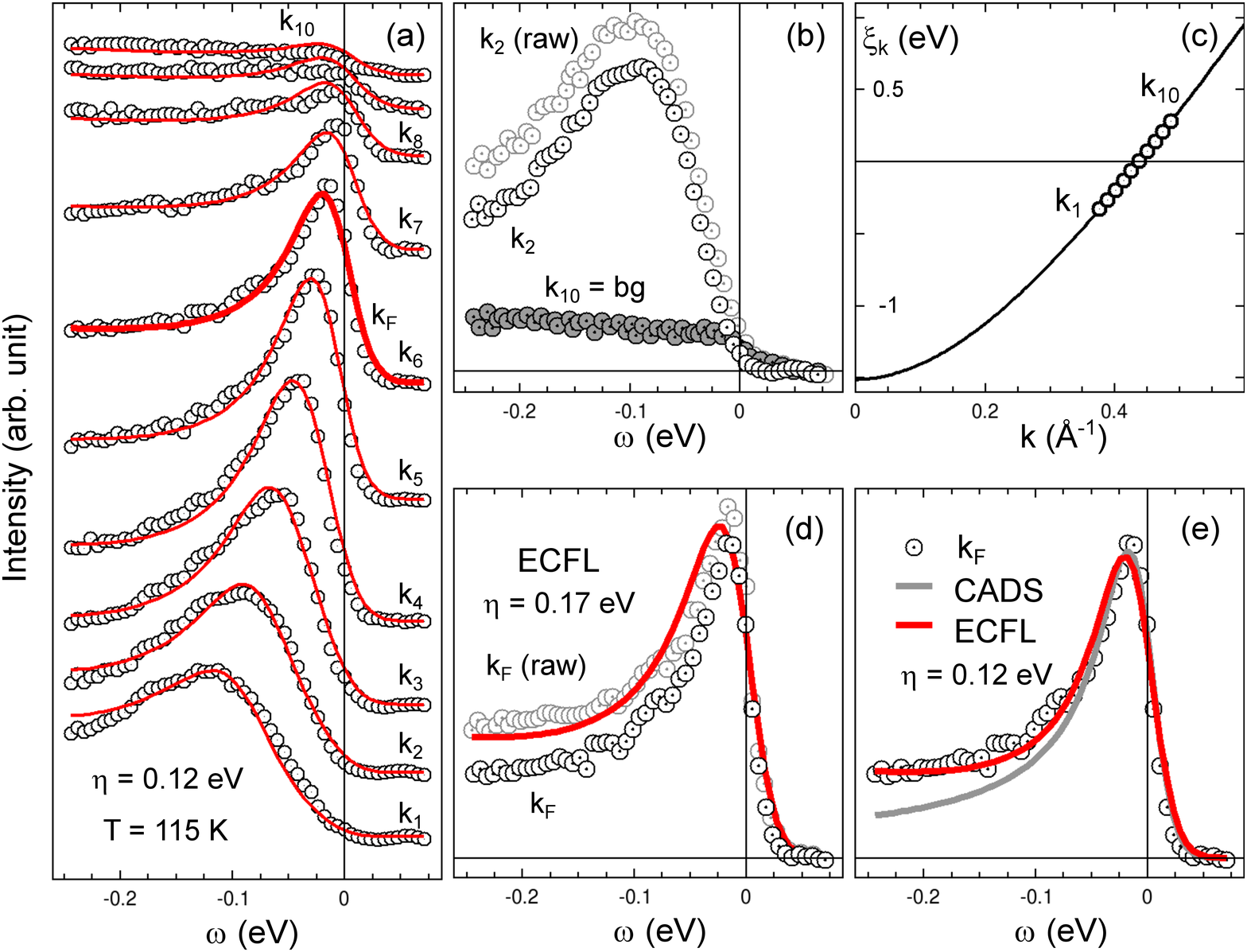}
	\caption{Conventional ARPES data (Bi2212) fit with the ECFL line
	shape.  The data are from Ref.\@
	\onlinecite{kaminski_renormalization_2001} ($T_c = 90$ K)\@.  (a) The
	data (symbols) and the fit (red lines) are shifted vertically by the
    same amount for ease of view.  (b)  An example of the raw data and the
    fit data is shown for $k_2$.  The background (bg) spectrum (see text)
    was subtracted from each raw data, and the resulting data, shown in
    (a), are then fit.  (c) The fixed $\xi_{\kv}$ parameters used for the
    fit.  Thus, in this figure, $\eta$ is the only fit parameter (cf.\@
    Fig.\@ \ref{fit-casey} caption).  (d) Raw data at $k=k_F$ fit with a
    somewhat greater $\eta$ value.  (e) The current fit compared with a fit
    using the CADS line shape.  \label{fit-kaminski}
	}
\end{center}
\end{figure}

{\bf Line shape fit for conventional ARPES:}\ \ We find that the magnitude
of the parameter $\omega_0$ ($0.5$ eV) determined from the fit of the sharp
laser data works very well also for the conventional ARPES data
\footnote{When it is allowed to vary but $k$-independent, it is 0.5 eV
within $\pm \sim$ 10 \%.}.  Thus, all parameters other than $\eta$ are
fixed, with one small exception in Fig.\@ \ref{fit-ours}(d), where a slight
change in $\omega_0$ produces a much better fit over a larger energy range
for LSCO\@.

Fig.\@ \ref{fit-kaminski} shows our fit of the data in Ref.\@
\onlinecite{kaminski_renormalization_2001} with a single free parameter
$\eta$.  The amount of the ``extrinsic background'' (bg) in ARPES is an
issue of importance \cite{olson_high-resolution_1990,
gweon_generalized_2003, kaminski_identifying_2004}, especially when
analyzing the conventional ARPES data.  Here we fit the bg
subtracted data, as well as the raw data (panel d).  For subtracting the
bg, we use an often-used procedure \cite{kaminski_identifying_2004,
valla_evidence_1999} of equating the background to a fraction (``bg scaling
factor'') of the data far beyond the Fermi surface crossing ($k = k_{10}$
for this data set).  The bg scaling factor, 1/2 for this figure, is
determined to be the maximum value for which the resulting intensity is not
negative.  As shown in the panel d, the ECFL fit remains good by adjusting
$\eta$, whether or not the extrinsic background is subtracted.  In
contrast, we find that the CADS theory, notwithstanding its notable
successes \cite{anderson_hidden_2008, casey_accurate_2008},  cannot cope
with even the background subtracted data (Fig.~3e), giving too steep a fall
off towards the left.  Likewise, the MFL fits
\cite{varma_phenomenology_1989, abrahams_what_2000} have been shown to
compare well with the data only after substantial background subtraction
\cite{valla_evidence_1999, kaminski_momentum_2005}.

\begin{figure}
\begin{center}
	\includegraphics[width=0.93\columnwidth]{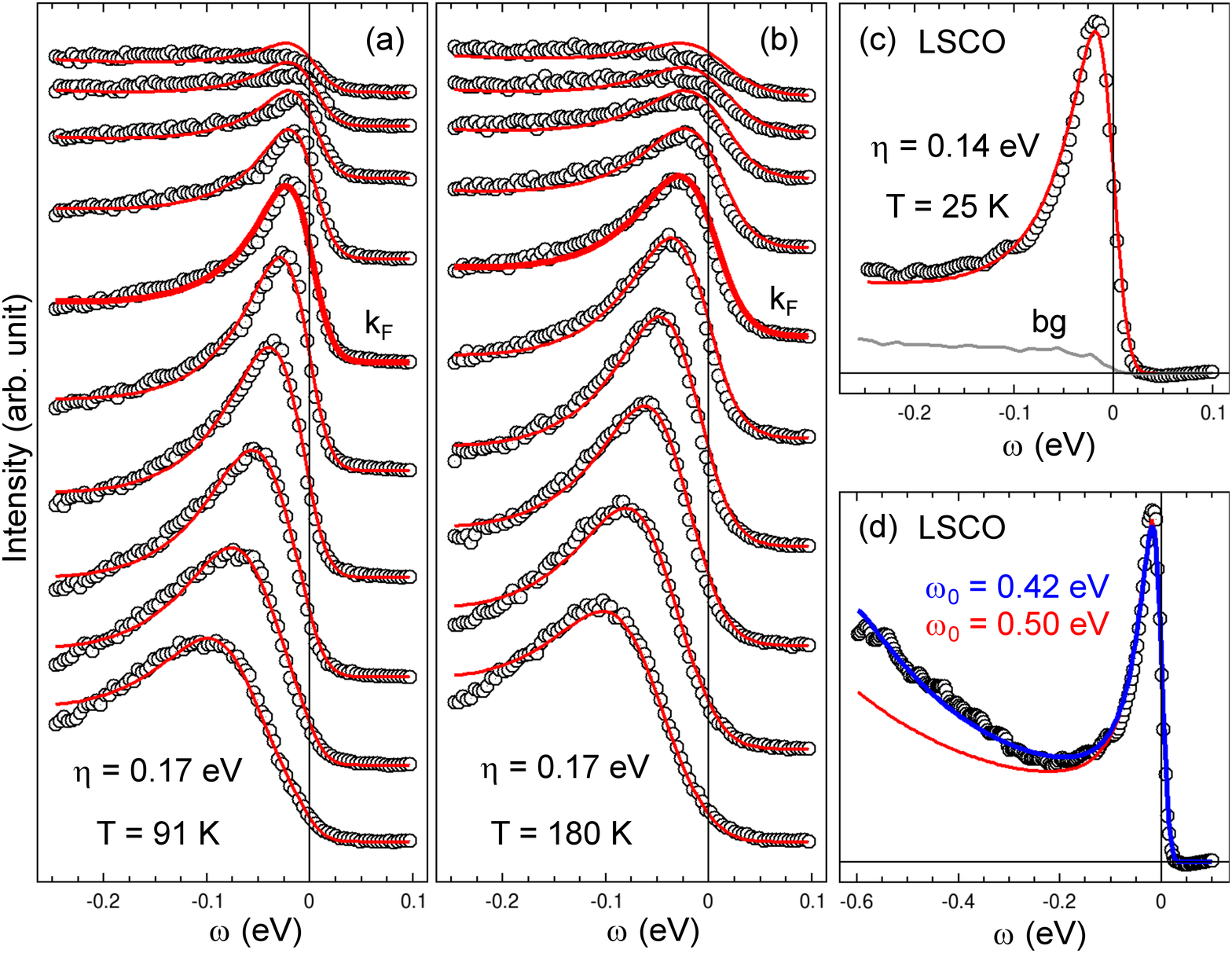}
	\caption{Conventional ARPES data, including our own (a,b), fit with the
	ECFL line shape.  The procedure used to fit these data are identical
	with those of the previous figure, i.e.\@ a fit with a single free
	parameter $\eta$, with (d) being a single exception.  (a, b) Optimally
	doped Bi2212 ($T_c = 91$ K).  (c) Optimally doped LSCO data
	\cite{yoshida_low-energy_2007, yoshida__2001}.  (d) A test fit up to
	0.6 eV for the LSCO data with a small change to $\omega_0$ for the same
	data  as in  (c) but over a wider energy range.  By changing $\omega_0$
	slightly from 0.50 eV to 0.42 eV, we see that an excellent fit up to
	0.6 eV is found.  The LSCO data, as far as we are aware, is fit only by
	the ECFL theory, since an energy dependence  rising linearly for
	occupied states occurs naturally and uniquely in the ECFL spectral
	function.
	\label{fit-ours}
	}
\end{center}
\end{figure}

Our own data on Bi2212 data, taken at $T_c$ and well above $T_c$, covering
a similar temperature range as the laser data of Fig.\@ \ref{fit-casey},
can be fit equally well with the same background subtraction procedure,
i.e.\@ with the  ``bg scaling factor'' (1/2), as shown in Fig.\@
\ref{fit-ours}.

We also find that the data for a lower-$T_c$ cuprate LSCO can be fit very
well with the same intrinsic parameters.  Here, we shall discuss only the
$k=k_F$ data for brevity.  In this case, we determine that the ``bg scaling
factor'' be 1.  The subtracted   ``bg'' data \cite{yoshida_low-energy_2007,
yoshida__2001} is shown as the  gray curve in Fig.\@ \ref{fit-ours}(c).
Given their weak superconductivity features \cite{yoshida_low-energy_2007,
yoshida__2001}, these LSCO data are taken to represent the normal state
property even if the temperature is slightly lower than $T_c$.  As for the
Bi2212 case, the data can be fit well even without the background
subtraction, if a somewhat greater $\eta$ value ($\approx 0.17$ eV) is
used.  It is clear, from Fig.\@ \ref{fit-ours}(c), that the data at a
temperature as low as 25 K can be fit very well with the ECFL line shape.
In addition, in working with LSCO line shapes, we noticed a steady and
rapid rise in intensity beyond $-\omega$ = 0.25 eV, a behavior different
from that of Bi2212.  We leave the full discussion of this non-universal
behavior for future work.  However, we find it exceptional that the current
theory is able to describe the line shape of LSCO up to very high energy,
as shown in Fig.\@ \ref{fit-ours}(d).

{\bf Kinks in the spectra:}\ \ The two independent energy scales $\omega_0$
and $\Delta_0$  are determined from our fit as $\sim 0.5$ eV and $\sim 0.1$
eV. These  are natural candidates for the two main dispersion anomalies in
the cuprates \cite{graf_universal_2007, lanzara_evidence_2001} as in Fig.\@
\ref{kinks}.  Well-defined energy distribution curve (EDC: intensity curve
at a fixed $\kv$ value) peaks disappear in a wide energy from $\sim 0.3$ eV
to $\sim$ 1 eV, as observed experimentally for the high energy kink
\cite{graf_universal_2007, meevasana_hierarchy_2007}.  As this feature
already exists in aux-FL, it cannot be associated with $\Delta_0$ but
rather with $\omega_0$. The (numerical) dynamical mean field theory
\cite{byczuk_kinks_2007} can already account for this feature as can  the
present ECFL (analytical) theory.

Turning to the low energy ARPES kink at $\sim 70$ meV,  Figs.\@
\ref{kinks}(c,d,e) illustrate the observed weak dispersion anomaly in the
normal state data (c), reproduced in the ECFL theory (d) but not in the
aux-FL theory (e).  Here we use a visualization method for momentum
distribution curve (MDC: intensity curve at a fixed $\omega$ value), an
object discussed primarily for low energy kinks.  Thus this feature
originates from the scale $\Delta_0$,   it causes an increased asymmetry
and the (blue)  shift of the peak to high hole energy, when the third term
in the caparison factor ($\frac{n^2}{4} \frac{\xi_{\kv} -
\omega}{\Delta_0}$) of Eq.\@ \ref{ecfl-A} becomes important.   To our
knowledge, the ECFL theory is a unique analytical theory that has both
these kink features arising from purely electronic (extreme) correlations.

\begin{figure}
\begin{center}
	\includegraphics[width=0.75\columnwidth]{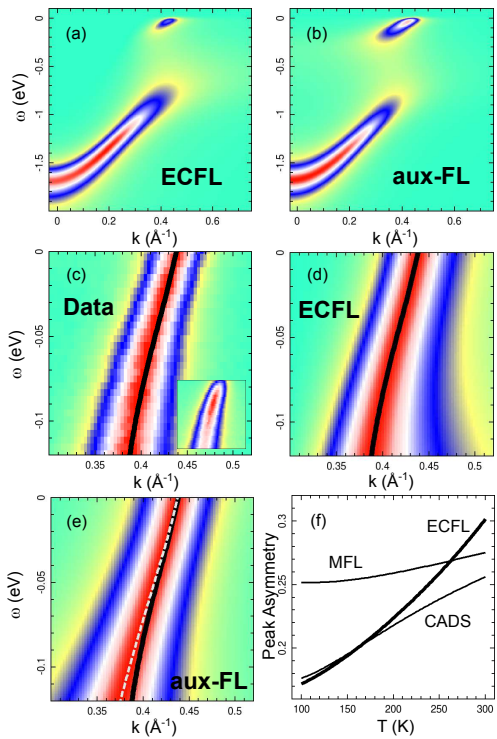}
	\caption{
	Image plots of the spectral function for (a) the  ECFL theory ($\eta
	= 0.17$ eV), and (b) that of the  auxiliary FL.  (c) The data of
	Fig.\@ \ref{fit-ours}(a) before (inset) and after (main)  MDC
	normalization, by which each MDC is scaled and shifted to have minimum
	0 (green) and maxmimum 1 (red).  Blue corresponds to 1/2.  (d) The
	near-$E_F$ part of the ECFL spectral function of (a), after
	MDC-normalization  with MDC peak positions traced by black line.  The
	black line in (c) is from (d).  (e) The near-$E_F$ part of the aux-FL
	spectral function of (b) after MDC-normalization.  The MDC peak
	positions are traced by gray dashed line, while the black line is from
	(d).  The bending such as shown by the black line here is commonly
	referred to as the kink.  (f) The temperature dependence of the peak
	asymmetry compared for three different theories, rising as $T^2$ for
	ECFL\@.  Theory parameters for the calculation, apart from $T$, are
	taken from the fit of Fig.\@ \ref{fit-casey}(a) for the ECFL, from the
	equivalent fit of Ref.\@ \onlinecite{casey_accurate_2008} for the CADS,
	and from Ref.\@ \onlinecite{kaminski_momentum_2005} for the MFL\@.
	\label{kinks}
	}
\end{center}
\end{figure}

In Fig.\@ \ref{kinks}(f), we show the temperature dependence of the
dimensionless peak skew or asymmetry, defined as (HL - HR) / (HL + HR),
where HR (HL) is the half-width at half maximum on the right (left) side of
the peak.  The predicted $T$-dependent asymmetry, predicted even greater
for $\eta \approx 0.15$ eV (synchrotron data; not shown), would be
interesting to explore in the future.

Further work is necessary to refine the picture suggested in this Letter.
For example, as $-\xi_{\kv}$ increases, the line shape becomes somewhat too
asymmetric.   Work is also in progress to apply the theory to two particle
response as seen, e.g.\@, in  optical conductivity.  We have checked that
the bubble approximation (conductivity as a product of two $G$'s) shows an
agreement in the order of magnitude of the frequency scale and the
conductivity.

{\bf Conclusions:}\ \ We have shown that it is possible to understand both
ARPES data sets (laser or conventional) comprehensively,  with identical
physical parameters.  Work going beyond the nodal cut and the optimal
doping value is in progress.  The theory is very tolerant of the
uncertainty in the background subtraction for the conventional ARPES data.
Additionally, the theory satisfies the global particle sum rule, and
contains two inter-dependent energy scales ($\omega_0$ and $\Delta_0$) that
correspond well to the energy scales of the two kinks.  Thus the simplest
version of the ECFL theory using  a small number of  parameters,  provides
a framework to understand the ARPES line shape  data  for the normal state
of the cuprates: {\em it works extremely well across techniques, samples
and temperatures}.

Portions of this research were carried out at the SSRL, a Directorate of
SLAC National Accelerator Laboratory and an Office of Science User Facility
operated for the U.S. DOE Office of Science by Stanford University, and at
the BNL, supported by the U.S. DOE under Grant No. DE-AC02-98CH10886.  The
work by GHG was supported partially by COR-FRG at UC Santa Cruz.  BSS was
supported by U.S. DOE under Grant No. FG02-06ER46319.

\end{document}